\documentstyle[preprint,aps]{revtex}

\newcommand{\itPhi}{\mbox{$\it\Phi$}}
\newcommand{\itGamma}{\mbox{$\it\Gamma$}}

\begin{document}

\draft
\title{Topological susceptibility at zero and finite temperature\\
       in the Nambu--Jona-Lasinio model}
\author{K. Fukushima\thanks{E-mail: fuku@nt1.c.u-tokyo.ac.jp},
        K. Ohnishi\thanks{E-mail: konishi@nt1.c.u-tokyo.ac.jp},
    and K. Ohta\thanks{E-mail: ohta@nt1.c.u-tokyo.ac.jp}}
\address{Institute of Physics, University of Tokyo,
3-8-1 Komaba, Meguro-ku, Tokyo 153-8902, Japan}

\maketitle

\begin{abstract}
We consider the three flavor Nambu--Jona-Lasinio model with the
't~Hooft interaction incorporating the ${\rm U}(1)_{\text{A}}$
anomaly. In order to set the coupling strength of the 't~Hooft term,
we employ the topological susceptibility $\chi$ instead of the $\eta'$
meson mass. The value for $\chi$ is taken from lattice simulations. We
also calculate $\chi$ at finite temperature within the model.
Comparing it with the lattice data, we extract information about the
behavior of the ${\rm U}(1)_{\text{A}}$ anomaly at finite
temperature. We conclude that within the present framework, the
effective restoration of the ${\rm U}(1)_{\text{A}}$ symmetry does not 
necessarily take place even at high temperature where the chiral
symmetry is restored.
\end{abstract}

\pacs{11.30.Rd, 12.39.Fe}

\section{Introduction}
The ${\rm U}(1)_{\text{A}}$ anomaly of QCD plays an essential role in
hadron physics. One of its most striking manifestations would probably
be the $\eta'$ meson mass. Since the ${\rm U}(1)_{\text{A}}$ symmetry
is broken not spontaneously but explicitly by the anomaly, $\eta'$
cannot be regarded as a nearly massless Nambu-Goldstone boson like the
other psuedoscalar mesons. In fact, the mass of $\eta'$ is as large as
the nucleon mass, i.e. $m_{\eta'}=958\  {\rm MeV}$. This is the
so-called ${\rm U}(1)_{\text{A}}$ problem.

The topological susceptibility, $\chi$, is an essential quantity in
considering the ${\rm U}(1)_{\text{A}}$ problem because it is related
to $m_{\eta'}$ through the Witten-Veneziano mass formula
\cite{witt,vene},
\begin{equation}
\frac{2N_{\text{f}}}{f_{\pi}^2}\chi=m_{\eta}^2+m_{\eta'}^2-2m_{K}^2,
\label{wv}
\end{equation}
where $N_{\text{f}}=3$ is the number of the flavors and $f_\pi$ is the
pion decay constant. This formula has been confirmed by calculating
$\chi$ directly on the lattice\cite{alle}: The calculations give
$\chi\sim (180\ {\rm MeV})^4$, which is consistent with the value
obtained by plunging into the formula (\ref{wv}) experimental values
of the pion decay constant and the meson masses. Thus, the topological
susceptibility could tell us as much information about the
${\rm U}(1)_{\text{A}}$ anomaly as does $m_{\eta'}$.

The tool we will employ in this work for the investigation of the
${\rm U}(1)_{\text{A}}$ problem is the three flavor
Nambu--Jona-Lasinio (NJL) model\cite{klev,hk,vogl,lehm} which can be
used as an effective theory of QCD. The NJL Lagrangian we adopt here
is
\begin{eqnarray}
\label{njl}
{\cal L}&=&{\cal L}_0+{\cal L}_4+{\cal L}_6,\\
{\cal L}_0&=&\bar{q}(i\gamma_\mu \partial^\mu-m)q,\\
{\cal L}_4&=&G\sum_{a=0}^8\left[(\bar{q}\lambda^a q)^2+
 (\bar{q}i\gamma_5\lambda^a q)^2\right],\\
{\cal L}_6&=&-K\left[\det\bar{q}(1+\gamma_5)q+
 \det\bar{q}(1-\gamma_5)q\right],
\label{thooft}
\end{eqnarray}
where the quark field $q$ is a column vector in the color, flavor and
spinor spaces, and $\lambda^{a}$ is the Gell-Mann matrix in the flavor
space with $\lambda^0=\sqrt{2/3}\;\text{diag}(1,1,1)$. The
determinants in Eq.\ (\ref{thooft}) are with respect to the flavor
indices.

The free quark Lagrangian ${\cal L}_0$ contains the current quark mass
term with $m=\text{diag}(m_{\text{u}},m_{\text{d}},m_{\text{s}})$,
breaking the ${\rm U}(3)_{\text{L}}\otimes{\rm U}(3)_{\text{R}}$
symmetry explicitly. Throughout this article, we assume the exact
isospin symmetry, i.e. $m_{\text{u}}=m_{\text{d}}$. The term ${\cal
L}_4$ generates the four-point couplings and is invariant under the
${\rm U}(3)_{\text{L}}\otimes{\rm U}(3)_{\text{R}}$ transformation.
The six-point, determinant term ${\cal L}_6$ is  what is called the
't~Hooft interaction and breaks the ${\rm U}(1)_{\text{A}}$ symmetry.
This interaction simulates the ${\rm U}(1)_{\text{A}}$ anomaly in our
scheme, and the effective coupling constant $K$ measures its
strength.

Let us review the status of parameter setting in the NJL model. The
parameters to be fixed are, the current quark masses
($m_{\text{u}}=m_{\text{d}}, m_{\text{s}}$), the three-momentum
cut-off ($\Lambda$), and the effective coupling constants ($G$ and
$K$). The physical quantities usually taken as inputs are, $m_\pi$,
$f_\pi$, $m_K$, and $m_{\eta'}$. The question we bring out here is as
to $m_{\eta'}$ which has been used for the determination of $K$. As is
well known, the NJL model lacks in confinement, and in fact, in this
model $\eta'$ decays into asymptotic $q\bar{q}$ states due to its
large mass. Thus $m_{\eta'}$ in the NJL model is not a well-defined
quantity. The alternative quantity for the determination of $K$ we
will use here is the topological susceptibility, which contains the
information about the ${\rm U}(1)_{\text{A}}$ anomaly and whose value
has been given by lattice Monte-Carlo simulations, as mentioned
above. One of the main purposes in the current work is to derive the
expression for the topological susceptibility within the framework of
the NJL model, and to fix the parameter $K$ by means of $\chi$ as an
input.

Recently, the behavior of the effect of the ${\rm U}(1)_{\text{A}}$
anomaly at finite temperature has been discussed intensively
\cite{shuryak,hatsuda,bernard,chan,janik}. In particular, special
attentions have been paid to whether or not the effective restoration
of the ${\rm U}(1)_{\text{A}}$ symmetry and the chiral phase
transition occur simultaneously. This question is still controversial
and is not settled yet. Here, we should clarify what we mean by `the
effective restoration of the ${\rm U}(1)_{\text{A}}$ symmetry.' It
means that all ${\rm U}(1)_{\text{A}}$ violating quantities vanish,
i.e.\ that all order parameters of the ${\rm U}(1)_{\text{A}}$
symmetry vanish.\footnote{The authors thank the unknown referee for
his suggestion on the definite meaning of the ${\rm U}(1)_{\text{A}}$
symmetry restoration.} The possibility that $\eta'$ would degenerate
with the other pseudoscalar mesons has a great deal of significance
upon the experimental view in relativistic heavy ion collisions
\cite{alko}. In the NJL model, the definition we gave above is
equivalent to $K$ getting to zero, since finite $K$ makes the
${\rm U}(1)_{\text{A}}$ symmetry breaking manifest in the NJL
lagrangian (\ref{njl}). Since the origin from which a finite value of
$K$ arises is the presence of instantons in the physical state, the
effective restoration of the ${\rm U}(1)_{\text{A}}$ symmetry is
expected owing to the naive argument that the instanton density will
be suppressed at sufficiently high temperature. Thus when the
magnitude of $K$ becomes smaller, we will call it `the effective
restoration of the ${\rm U}(1)_{\text{A}}$ symmetry.'

The temperature dependence of $K$ in the NJL model, which indicates
nothing but the temperature dependence of the ${\rm U}(1)_{\text{A}}$
anomaly, has been put by hand and not gone beyond phenomenology
\cite{hk}. This is both because experimental data for $m_{\eta'}$ at
finite temperature, which are necessary for determination of $K$, are
not available, and because $\eta'$ becomes unbound completely in the
model soon after we raise the temperature from zero. Since the
topological susceptibility has been calculated at finite temperature
on the lattice, we will be able to determine the temperature
dependence of $K$ using that data, and obtain some knowledge about the
effective restoration of the ${\rm U}(1)_{\text{A}}$ symmetry.

The paper is organized as follows. In Sec.\ \ref{sec2}, we will derive
the expression for $\chi$ in the NJL model. Sec.\ \ref{sec3} is
devoted to parameter setting of the model and numerical calculations
of the physical quantities. A summary is given in Sec.\ \ref{sec4}.

\section{Topological susceptibility in the NJL model}
\label{sec2}
In this section, we calculate the topological susceptibility, $\chi$,
within the framework of the NJL model.

The first task is to know a general expression of $\chi$. We
recapitulate here the definition of $\chi$. We begin with the QCD
Lagrangian density
\begin{equation}
{\cal L}_{\text{QCD}}=-\frac{1}{4}F_{\mu\nu}^a F^{a\mu\nu}+
 \bar{q}(i\gamma_\mu D^\mu-m)q+\theta Q,
\end{equation}
where $F_{\mu\nu}^a$ is the gluon field strength tensor,
$D_\mu=\partial_\mu+igA_\mu$ is the covariant derivative with $A_\mu$
being the gluon field, $\theta$ is the QCD vacuum angle, and $Q$ is
the topological charge density defined by
\begin{equation}
Q(x)=\frac{g^2}{32\pi^2}F_{\mu\nu}^a\tilde{F}^{a\mu\nu}.
\end{equation}
With this Lagrangian density, the vacuum energy density $\varepsilon$
is written as
\begin{equation}
e^{-\varepsilon VT}=\int{\cal D}A_\mu{\cal D}\bar{q}{\cal D}q
 \,e^{\int d^4 x{\cal L}_{\text{QCD}}} \equiv Z
\end{equation}
in a path integral form, where $V$ and $T$ are the space and time
volumes, respectively. The topological susceptibility $\chi$ is
defined as a second derivative of $\varepsilon$ with respect to
$\theta$ at $\theta =0$,
\begin{equation}
\chi\equiv\left.\frac{\partial^2\varepsilon}{\partial\theta^2}
 \right|_{\theta =0}=\int d^4 x \langle 0|{\rm T}Q(x)Q(0)|0
 \rangle _{\text{connected}},
\end{equation}
where T stands for the time ordering operator, and the subscript
`connected' means to pick out the diagrammatically connected
contributions. Thus, in order to calculate $\chi$ in the NJL model, it
is necessary to find a correspondent to $Q(x)$ in the model. For that
purpose, we consider the four-divergence of the
${\rm U}(1)_{\text{A}}$ current,
$J_{5\mu}=\bar{q}\gamma_\mu \gamma_5 q$. In QCD, one has 
\begin{equation}
\partial^\mu J_{5\mu}=2N_{\text{f}}Q(x)+2i\bar{q}m\gamma_5 q,
\end{equation}
which does not vanish due to the anomaly.

On the other hand, in the NJL model
(Eqs.\ (\ref{njl})$\sim$(\ref{thooft})), we find\cite{hk}
\begin{equation}
\partial^\mu J_{5\mu}=4N_{\text{f}}K\text{Im}\det\itPhi
 +2i\bar{q}m\gamma_5 q,
\end{equation}
where
\begin{equation}
\itPhi_{ij}=\bar{q}_{i}(1-\gamma_{5})q_{j},
\end{equation}
and $i,j$ denote the flavor indices. Comparing these two expressions,
we find that
\begin{eqnarray}
Q(x) &=& 2K\text{Im}\det\itPhi \nonumber\\
&=&-iK\left[\det\itPhi-\left(\det\itPhi\right)^* \right]
\end{eqnarray}
in the NJL model.

With the definition of $\itGamma_\pm \equiv 1\pm\gamma_5$, we can
write
\begin{eqnarray}
\det\itPhi &=& \frac{1}{3!}\epsilon^{abc}\epsilon^{ijk}
 (\bar{q}_i \itGamma_- q_a)(\bar{q}_j \itGamma_- q_b)
 (\bar{q}_k \itGamma_- q_c),\nonumber\\
(\det\itPhi)^* &=& \frac{1}{3!}\epsilon^{def}\epsilon^{lmn}
 (\bar{q}_l\itGamma_+ q_d)(\bar{q}_m \itGamma_+ q_e)
 (\bar{q}_n \itGamma_+ q_f)
\end{eqnarray}
so that
\begin{eqnarray}
\chi &=& \int d^4 x \langle 0|{\rm T}Q(x)Q(0)|0
 \rangle_{\text{connected}} \nonumber\\
&=&-\frac{K^2}{(3!)^2}\int d^4 x\,\epsilon^{abc}\epsilon^{ijk}
 \epsilon^{def}\epsilon^{lmn} \nonumber\\
\times\langle 0|&{\rm T}&\left\{
 (\bar{q}_i \itGamma_- q_a)(\bar{q}_j \itGamma_- q_b)
 (\bar{q}_k \itGamma_- q_c)(x)(\bar{q}_l \itGamma_- q_d)
 (\bar{q}_m \itGamma_- q_e)(\bar{q}_n \itGamma_- q_f)(0)\right.
 \nonumber\\
&&-(\bar{q}_i \itGamma_- q_a)(\bar{q}_j \itGamma_- q_b)
 (\bar{q}_k \itGamma_- q_c)(x)(\bar{q}_l \itGamma_+ q_d)
 (\bar{q}_m \itGamma_+ q_e)(\bar{q}_n \itGamma_+ q_f)(0)
 \nonumber\\
&&-(\bar{q}_i \itGamma_+ q_a)(\bar{q}_j \itGamma_+ q_b)
 (\bar{q}_k \itGamma_+ q_c)(x)(\bar{q}_l \itGamma_- q_d)
 (\bar{q}_m \itGamma_- q_e)(\bar{q}_n \itGamma_- q_f)(0)
 \nonumber\\
&&+(\bar{q}_i \itGamma_+ q_a)(\bar{q}_j \itGamma_+ q_b)
 (\bar{q}_k \itGamma_+ q_c)(x)(\bar{q}_l \itGamma_+ q_d)
 (\bar{q}_m \itGamma_+ q_e)(\bar{q}_n \itGamma_+ q_f)(0)
 \left.\right\}|0\rangle_{\text{connected}}.
\label{chi1}
\end{eqnarray}
Now we must evaluate these four matrix elements. For the time being,
we pick up one term out of the four. Following Wick's theorem, we take
full contraction in terms of the propagator $S(x,x')$ that has been
constructed in the self-consistent gap equation
\cite{klev,hk,vogl,lehm}. Although several ways of contraction are
possible, there exists only one that is the leading order in expansion
in terms of inverse powers of the number of colors, i.e.
$1/N_{\text{c}}$. The situation is demonstrated in Fig.\ \ref{fig1} by 
means of the finite range representation.

The diagrams (a) and (b) in Fig.\ \ref{fig1} contain
${N_{\text{c}}}^4$ coming from traces over color. The diagram (c)
contains ${N_{\text{c}}}^5$ and is the leading order in
$1/N_{\text{c}}$ expansion. Notice that (b) is the exchange term for
(c) and is lowered by $1/N_{\text{c}}$ as compared with (c). Since the
gap equation for the constituent quark masses and the dispersion
equations for the meson masses are derived up to the leading order of
the large $N_{\text{c}}$ expansion \cite{klev}, we should take only
the contribution of (c) for the consistent treatment.

Taking account of the four terms in Eq.\ (\ref{chi1}), we obtain the
following expression for the lowest order of the diagrammatical
expansion,
\begin{eqnarray}
\chi^{\text{(lowest)}} &=& -\frac{K^2}{(3!)^2}(-9)\,\epsilon^{abc}
 \epsilon^{ijk}\epsilon^{def}\epsilon^{lmn}\,4\left\{\int d^4 x
 N_{\text{c}}\text{tr}\left[S_{di}(x)\gamma_5 S_{al}(x)\gamma_5
 \right]\right\} \nonumber\\
&&\times{N_{\text{c}}}^4\text{tr}\left[S_{bj}(0)\right]\text{tr}
 \left[S_{ck}(0)\right]\text{tr}\left[S_{em}(0)\right]\text{tr}
 \left[S_{fn}(0)\right],
\label{chi2}
\end{eqnarray}
where the full propagator in the Euclidean space is given as
\begin{equation}
S_{ij}(x)=\delta_{ij}\int\frac{d^4 p}{(2\pi)^4}\frac{
\ooalign{\hfil/\hfil\crcr$p$}+m^*_i}{-p^2-{m^*_i}^2}e^{-ip\cdot x},
\end{equation}
and $m_i^*$ denotes the constituent quark mass.

The object in the curly brackets in Eq.\ (\ref{chi2}) corresponds to
the one-loop part connecting the points $x$ and $0$ in Fig.\
\ref{fig1}(c). This is basically the one-loop proper polarization
insertion $\Pi_{ij}^{\text{ps}}(k^2)$ \cite{klev,hk,vogl,lehm} with
$k^2=0$ although the trace over flavor is not taken in this case.
Especially in the case of $a=i$ and $d=l$, which is the condition for
giving non-zero contribution to $\chi^{(lowest)}$ due to the
$\varepsilon$-tensor
in Eq.\ (\ref{chi2}), the object can be identified with an element of
3-, 8-, 0-channel polarizations with $k^2=0$; for instance, 0-0
channel polarization is
\begin{eqnarray}
\Pi_{00}^{\text{ps}}(k^2=0)&=&\text{tr}_{\text{flavor}}\frac{2}{3}
 \>\text{diag}\left(-N_{\text{c}}\int\frac{d^4 p}{(2\pi)^4}\text{tr}
 \left[i\gamma_5 S^{\text{u}}(p)i\gamma_5 S^{\text{u}}(p)\right],
 \right.\nonumber\\
&&\left.-N_{\text{c}}\int\frac{d^4 p}{(2\pi)^4}\text{tr}
 \left[i\gamma_5 S^{\text{d}}(p)i\gamma_5 S^{\text{d}}(p)\right],
 -N_{\text{c}}\int\frac{d^4 p}{(2\pi)^4}\text{tr}\left[i\gamma_5
 S^{\text{s}}(p)i\gamma_5 S^{\text{s}}(p)\right]\right).
\end{eqnarray}

Actually, there exist other diagrams that are of the same order in
$1/N_{\text{c}}$ expansion as the diagram in Fig.\ \ref{fig1}(c). They
are shown in Fig.\ \ref{fig2}. They are of the same order as
Fig.\ \ref{fig1}(c) because, while each four-point vertex is of
$O({N_{\text{c}}}^{-1})$\cite{klev}, it is compensated by a factor
$N_{\text{c}}$ coming from its neighboring loop. We have to include
all these diagrams for consistency of the $1/N_{\text{c}}$ expansion.
We will call their sum $\chi^{\text{(ring)}}$, for they are regarded
as the ring diagrams to be resummed in the mean field approximation.

Of course, the sum of these ring diagrams with the one-loop diagram
included can be interpreted as a propagation of a certain meson. Note
that the momentum of the propagating particle is zero; $k^{2}=0$. This
is just the reflection of the fact that $\chi$ is the quantity of the
zero frequency mode of the Fourier transform of
$\langle\text{T}Q(x)Q(0)\rangle $, namely,
\begin{equation}
\chi =\left.\int d^4 x\,e^{-ik\cdot x}\langle{\rm T}Q(x)Q(0)\rangle
 \right|_{k=0}.
\end{equation}

Now we calculate $\chi^{\text{(ring)}}$ following the Feynman rules of
the NJL Lagrangian. We note that the diagrams in Fig.\ \ref{fig2} are
obtained by replacing the one-loop polarization part in
Fig.\ \ref{fig1}(c) with the ring diagrams. Correspondingly we can
obtain $\chi^{\text{(ring)}}$ by replacing the curly bracketed part in
Eq.\ (\ref{chi2}) with the sum of the ring diagrams. We first perform
the summation over flavor indices. After that, it can be shown that
the expressions corresponding to each edge point of a diagram in
Fig.\ \ref{fig2} are brought together into a matrix form and are
arranged to the linear combination of $\lambda_8$ and $\lambda_0$
matrices. In other words, the 8- and 0-channel vertices have been
assigned to each edge point of each ring diagram in Fig.\ \ref{fig2}.
Then following the Feynman rules, we can construct the ring diagrams
by linking 8-, 0-channel polarizations
$\Pi_{88}^{\text{ps}}(k^2\!=\!0)$,
$\Pi_{80}^{\text{ps}}(k^2\!=\!0)=\Pi_{08}^{\text{ps}}(k^2\!=\!0)$,
$\Pi_{00}^{\text{ps}}(k^2\!=\!0)$, and 8-, 0-channel vertices
$K_{88}^{(+)}$, $K_{80}^{(+)}=K_{08}^{(+)}$, $K_{00}^{(+)}$, in all
possible ways. The result is
\begin{eqnarray}
\chi^{\text{(ring)}} &=& 4{N_{\text{c}}}^4 K^2\Bigg\{\frac{1}
 {\sqrt{3}}\text{tr}S^{\text{u}}\left(\text{tr}S^{\text{s}}-\text{tr}
 S^{\text{u}}\right){\Pi_{88}\choose \Pi_{80}}^{\text{t}}+\frac{1}
 {\sqrt{6}}\text{tr}S^{\text{u}}\left(2\text{tr}S^{\text{s}}
 +\text{tr}S^{\text{u}}\right){\Pi_{08}\choose \Pi_{00}}^{\text{t}}
 \Bigg\} \nonumber\\
&&\times\;2\hat{K}\left(1-2\hat{\Pi} \hat{K}\right)^{-1} \nonumber\\
&&\times \Bigg\{\frac{1}{\sqrt{3}}\text{tr}S^{\text{u}}\left(
 \text{tr}S^{\text{s}}-\text{tr}S^{\text{u}}\right)
 {\Pi_{88}\choose \Pi_{08}}+\frac{1}{\sqrt{6}}\text{tr}S^{\text{u}}
 \left(2\text{tr}S^{\text{s}}+\text{tr}S^{\text{u}}\right)
 {\Pi_{80}\choose \Pi_{00}}\Bigg\},
\label{chi'}
\end{eqnarray}
where
\begin{equation}
\hat{K} = \Bigg(
 \begin{array}{cc}
  K_{88}^{(+)} & K_{80}^{(+)}\\
  K_{08}^{(+)} & K_{00}^{(+)}
 \end{array}
\Bigg),\qquad
\hat{\Pi} = \Bigg(
 \begin{array}{cc}
 \Pi_{88} & \Pi_{80}\\
 \Pi_{08} & \Pi_{00}
 \end{array}
\Bigg)
\end{equation}
are $2\times 2$ matrices, and
\begin{equation}
 \Pi_{ij}=\Pi_{ij}^{\text{ps}}(k^2=0),\qquad
 \text{tr}S^{i}=\text{tr}S^{i}(x=0).
\end{equation}

For example, the first term in the first curly brackets of
Eq.\ (\ref{chi'}) produces contributions from the diagrams whose one
edge point is an 8-channel vertex. And, again for example, the second
term in the second curly brackets of Eq.\ (20) produces contributions
from the diagrams the other edge point of which is a 0-channel vertex.
The 8-8 and 0-0 channel diagrams can be interpreted as the
propagations of $\eta_{8}$ and $\eta_{0}$ mesons respectively. We see
that in addition to the $\eta_{8}$ and $\eta_{0}$ propagations, there
occur 8-0 and 0-8 mixing channel diagrams.

Another comment is in order. If
$\text{tr}S^{\text{u}}=\text{tr}S^{\text{s}}$, that is, the
${\rm SU}(3)_{\text{V}}$ symmetry is exact, the 8-channel vertices in
Eq.\ (\ref{chi'}) vanish. In this case, 8- and mixing channel
polarizations $\Pi_{88}^{\text{ps}}$,
$\Pi_{80}^{\text{ps}}=\Pi_{08}^{\text{ps}}$ as well as 8- and mixing
channel vertices $K_{88}^{(+)}$, $K_{80}^{(+)}=K_{08}^{(+)}$ all
vanish, so that only the ring diagrams constructed by those of the
0-channel, $\Pi_{00}^{\text{ps}}$ and $K_{00}^{(+)}$, contribute to
$\chi^{\text{(ring)}}$, which are interpreted as the propagation of
the pure $\eta_0$ state.

Finally, combining Eqs.\ (\ref{chi2}) and (\ref{chi'}), we arrive at
the expression for the topological susceptibility,
\begin{eqnarray}
\chi &=& 4{N_{\text{c}}}^4 K^2\Bigg[-N_{\text{c}}\left(
 \text{tr}S^{\text{u}}\right)^4 \left(\text{tr}S^{\text{s}}\right)^2
 \left(\frac{2}{m_{\text{u}}^{*}\text{tr}S^{\text{u}}}+
 \frac{1}{m_{\text{s}}^{*}\text{tr}S^{\text{s}}}\right) \nonumber\\
&& +\Bigg\{\frac{1}{\sqrt{3}}\text{tr}S^{\text{u}}\left(\text{tr}
 S^{\text{s}}-\text{tr}S^{\text{u}}\right)
 {\Pi_{88}\choose \Pi_{80}}^{\text{t}}+\frac{1}{\sqrt{6}}\text{tr}
 S^{\text{u}}\left(2\text{tr}S^{\text{s}}+\text{tr}S^{\text{u}}
 \right){\Pi_{08}\choose \Pi_{00}}^{\text{t}}\Bigg\} \nonumber\\
&&\times\;2\hat{K}\left(1-2\hat{\Pi} \hat{K}\right)^{-1} \nonumber\\
&&\times \Bigg\{\frac{1}{\sqrt{3}}\text{tr}S^{\text{u}}
 \left(\text{tr}S^{\text{s}}-\text{tr}S^{\text{u}}\right)
 {\Pi_{88}\choose\Pi_{08}}+\frac{1}{\sqrt{6}}\text{tr}S^{\text{u}}
 \left(2\text{tr}S^{\text{s}}+\text{tr}S^{\text{u}}\right)
 {\Pi_{80}\choose\Pi_{00}}\Bigg\}\Bigg].
\end{eqnarray}

Let us give one more comment. In general, a two-point correlation
function of gauge invariant operators may be decomposed into the sum
over multi-particle intermediate states by inserting a complete set
between two operators. It is known that the dominant contributions of
the leading order in $1/N_{\text{c}}$ expansion are those of
one-particle intermediate states. Moreover Witten\cite{witt} and
Veneziano\cite{vene} have derived their formula by assuming that, when
the momentum of the intermediate particle is zero, which is the case
of $\chi$, the contribution of the $\eta_0$ propagating state is the
only leading term in $1/N_{\text{c}}$ expansion. These statements are
quite consistent with our specific model calculation respecting large
$N_{\text{c}}$ expansion (although $\eta_8$ and mixing channels
besides $\eta_0$ propagate in the intermediate states in our model due
to the explicit ${\rm SU}(3)_{\text{V}}$ symmetry breaking).

\section{Numerical calculation}
\label{sec3}
Now that we have obtained the expression for the topological
susceptibility, we proceed to numerical calculations. In
Sec.\ \ref{sub31}, we set the parameters at zero temperature by
employing $\chi$. With the determined parameters, we calculate
physical quantities. In Sec.\ \ref{sub32}, we discuss the temperature
dependence of $\chi$ and the six-point coupling constant $K$.

\subsection{Parameter setting at zero temperature}
\label{sub31}

The parameters to be set in the NJL model are;
\begin{eqnarray*}
\left\{
\begin{array}{lp{1.5cm}c}
 \mbox{current quark masses} &&
  m_{\text{u}}=m_{\text{d}}, \quad m_{\text{s}} \\
 \mbox{three-momentum cut-off} && \Lambda \\
 \mbox{four-point coupling constant} && G \\
 \mbox{six-point coupling constant} && K.
\end{array}
\right.
\end{eqnarray*}

As for $m_{\text{u}}=m_{\text{d}}$, we set them to be
$m_{\text{u}}=m_{\text{d}}=5.5\ {\rm MeV}$ following Ref.\ \cite{hk}.

To set the other four parameters, we use the following quantities as
inputs,
\begin{eqnarray*}
\left\{
\begin{array}{lcc}
 m_\pi &=& 138\ {\rm MeV} \\
 f_\pi &=& 93\ {\rm MeV} \\
 m_K &=& 495.7\ {\rm MeV} \\
 \chi &=& (175\pm5 {\rm MeV})^4.
\end{array}
\right.
\end{eqnarray*}
The fourth quantity we use here in place of
$m_{\eta'}=957.5\ {\rm MeV}$ is, as mentioned in the introduction, the
topological susceptibility $\chi$. The numerical value of $\chi$ is
taken from Ref.\ \cite{alle}, in which $\chi$ is calculated in the
quenched approximation.

Initially, however, we will calculate with the parameters determined
by using $m_{\eta'}$ as input in order to check consistency of
$m_{\eta'}$ and $\chi$ in the NJL model. Parameter setting with
$m_{\eta'}$ has been performed in Ref.\ \cite{hk}, and the results are
\begin{displaymath}
m_{\text{s}}=135.7\ {\rm MeV},\quad \Lambda=631.4\ {\rm MeV},
 \quad G\Lambda^2=1.835,\quad K\Lambda^5=9.29.
\end{displaymath}
The physical quantities calculated from these parameters are
summarized in the first column of Table \ref{table}. We first check
the Witten-Veneziano mass formula (\ref{wv}) within the NJL model. The
computed values of $\chi$ and $m_\eta$ with those parameters of
Ref.\ \cite{hk} are
\begin{eqnarray*}
\left\{
\begin{array}{ccc}
 \chi_{\text{NJL}} &=& (166\ {\rm MeV})^4 \\
 m_\eta &=& 487\ {\rm MeV},
\end{array}
\right.
\end{eqnarray*}
so that the ratio of LHS to RHS in Eq.\ (\ref{wv}) turns out to be
\begin{equation}
\frac{2N_{\text{f}}\chi}{f_\pi^2(m_\eta^2+m_{\eta'}^2-2m_K^2)}=0.81.
\end{equation}
On the other hand, the ratio of $\chi_{\text{NJL}}$ to
$\chi_{\text{Lat}}$, which means how much the conventional parameters
determined with $m_{\eta'}$ reproduces the lattice data of $\chi$, is
\begin{equation}
\frac{\chi_{\text{NJL}}}{\chi_{\text{Lat}}} = 0.80.
\end{equation}
>From the above two ratios, we can say that as a whole, $m_{\eta'}$ and
$\chi$ are reproduced well simultaneously in the NJL model.

Now we consider parameter setting with $\chi$ used. The topological
susceptibility, $\chi$ might be a more suitable quantity for parameter
setting than $m_{\eta'}$ for the following two reasons:
\begin{itemize}
\item Since $\eta'$ decays into asymptotic $q\bar{q}$ state due to
lack of confinement in the NJL model, $m_{\eta'}$ may be a less
reliable quantity, while $\chi$ is free from such a shortcoming of the
NJL model.
\item The value of $\chi$, $(175\ {\rm MeV})^4$ is small enough
compared with the cut-off $\Lambda\sim 600\ {\rm MeV}$. Thus the NJL
model is expected to describe $\chi$ well.
\end{itemize}

The parameters obtained by using $\chi=(175\ {\rm MeV})^4$ are
\begin{displaymath}
m_{\text{s}}=135.7\ {\rm MeV},\quad \Lambda=631.4\ {\rm MeV},\quad
 G\Lambda^2=1.765,\quad K\Lambda^5=11.32.
\end{displaymath}
We note that $K\Lambda^5$ becomes larger than the case of using
$m_{\eta'}$ as an input, which implies that the binding of $\eta'$ is
loosened. (The 't~Hooft interaction loosens the binding of mesons,
that is, induces a repulsive force between quarks. This can be seen
from the very fact that $\eta'$ becomes massive due to the
interaction.) Physical quantities calculated with these parameters are
shown in the second column of Table \ref{table}. The solution for
$m_{\eta'}$ in the mean field approach does not exist, that is,
$\eta'$ is not bound any more. We see that $m_{\eta}$ is improved
slightly. Although $\eta'$ no longer exists in the NJL model, we could
infer its mass by utilizing the Witten-Veneziano mass formula
(\ref{wv}); $m_{\eta'}=942\ {\rm MeV}$ is obtained.

We close this subsection by referring to the study due to Takizawa,
Nemoto, and Oka \cite{taki}, in which the parameters, especially the
six-point coupling constant $K$, are determined in a  different
approach. They examined the radiative decays of an $\eta$ meson such
as $\eta\rightarrow2\gamma$, $\eta\rightarrow\gamma l^- l^+$ and
$\eta\rightarrow\pi^0\gamma\gamma$, and obtained rather strong
six-point coupling constant $K\Lambda^5$, namely, four times as large
as that determined by using $m_{\eta'}$. Although we cannot compare
our parameters directly with theirs due to different cut-off schemes
(their scheme is the four-momentum cut-off), it is not probable that
our result is compatible with theirs. Still, we believe that our approach
is rather straightforward to probe the ${\rm U}(1)_{\text{A}}$ anomaly.

\subsection{Behavior of $K$ at finite temperature}
\label{sub32}
In this subsection, we discuss the temperature dependence of $K$,
comparing the NJL calculation of $\chi$ with the lattice data.

The lattice data for the topological susceptibility \cite{alle} are
shown in Fig.\ \ref{fig3} with error bars. The $T_{\text{c}}$ in the
figure denotes the temperature of the chiral phase transition.
Although $T_{\text{c}}=260\ {\rm MeV}$ in the original
Ref.\ \cite{alle}, we have rescaled it to $150\ {\rm MeV}$. We should
notice that the lattice data are computed only up to
$T=1.4T_{\text{c}}$. Unfortunately, the lattice data are absent at
high temperatures. At any rate, the data show that $\chi$ drops
rapidly around $T_{\text{c}}$.

One comment should be noted. The fact that $\chi$ drops near
$T_{\text{c}}$ does not always mean the effective restoration of the
${\rm U}(1)_{\text{A}}$ symmetry at $T_{\text{c}}$. This can be seen
by returning to the Witten-Veneziano mass formula (\ref{wv}),
\begin{equation}
2N_{\text{f}}\chi = f_\pi^2(m_\eta^2+m_{\eta'}^2-2m_K^2).
\eqnum{$1'$}
\end{equation}
We realize that the pion decay constant $f_\pi$, which is associated
with the spontaneous chiral symmetry breaking (${\rm SU}(3)_{\text{L}}
\otimes{\rm SU}(3)_{\text{R}}\rightarrow{\rm SU}(3)_{\text{V}}$), has
entered the formula. Since $f_\pi$ would become zero along with the
restoration of the chiral symmetry, $\chi$ is also expected to become
zero around $T_{\text{c}}$. In this sense, the lattice data which show
the dropping of $\chi$ at $T_{\text{c}}$ is what should be expected
from the formula, and rather, we could consider that the data confirm
the validity of the Witten-Veneziano mass formula. Thus the dropping
of $\chi$ in the lattice data should be attributed to the restoration
of the chiral symmetry, and does not always indicate the effective
restoration of the ${\rm U}(1)_{\text{A}}$ symmetry. It is worth
noting that this behavior results from large $N_{\text{c}}$ expansion.
In Ref.\ \cite{birse} it was pointed out that $n$-point correlation
functions ($n<N_{\text{f}}=3$) cannot detect any effect of the
${\rm U}(1)_{\text{A}}$ anomaly in the chiral symmetric phase. One
might have thought that the dropping of $\chi$ near $T_{\text{c}}$
would be regarded as $\chi$'s insensitivity to the
${\rm U}(1)_{\text{A}}$ anomaly. However, it is not the case because
$\chi$ is not a ${\rm U}(1)_{\text{A}}$ singlet quantity. In fact, it
contains contributions carrying the ${\rm U}(1)_{\text{A}}$ charge 2,
0 and -2. Thus $\chi$ is an appropriate quantity to observe the fate
of the ${\rm U}(1)_{\text{A}}$ symmetry even in the chiral symmetric
phase, in principle, even though careful attention should be paid in
order to infer correct meanings.

In this respect, the discussions of Schaffner-Bielich\cite{schaffner}
is obscure; he discusses under the assumption that $\chi$
and the ${\rm U}(1)_{\text{A}}$ anomaly are equivalent to each other
and that the dropping of $\chi$ at $T_{\text{c}}$ immediately means
the effective restoration of the ${\rm U}(1)_{\text{A}}$ symmetry. His
assumption is considered as correct only if the dropping rate of
$\chi$ is much faster than that of $f_\pi$. To judge the validity for
this prevailing assumption is what we pursue in the present work. In
fact, as discussed below, our result reveals that the assumption has
no convincing reliability, at least, within the framework of the NJL
model.

Now we consider the temperature dependence of $\chi$ in the NJL
model. Among the four parameters ($m_{\text{s}}, \Lambda, G, K$), we
might reasonably fix $m_{\text{s}}$ and $\Lambda$ at the values
determined at zero temperature. In general, however, we should take
account of temperature dependences of the coupling constants
$G\Lambda^2$ and $K\Lambda^5$. As for $G\Lambda^2$, it would be hard
or almost hopeless to get information about the temperature
dependence even in some phenomenological sense. Here, we make an
assumption that $G\Lambda^2$ does not depend on temperature. This
might be partially justified by the fact that even if $G\Lambda^2$ is
constant, the NJL model restores the chiral symmetry as a consequence
of its own dynamics.

We now pay attention to the behavior of $K\Lambda^5$, which indicates
the temperature dependence of the ${\rm U}(1)_{\text{A}}$ anomaly. For
the first case, we treat $K\Lambda^5$ as a constant parameter and fix
it at the values at zero temperature. We will call this prescription
CASE A. The calculated temperature dependence of $\chi$ is shown in
Fig.\ \ref{fig3} (the solid line). We see that $\chi$ in the NJL model
drops near $T_{\text{c}}$ as the dynamical consequence and reproduces
the lattice data up to $1.4T_{\text{c}}$ considerably well. This
result means that the ${\rm U}(1)_{\text{A}}$ symmetry is not restored
at least up to $1.4T_{\text{c}}$, and we conclude that the effective
restoration of the ${\rm U}(1)_{\text{A}}$ symmetry does not coincide
with the chiral phase transition.

At high temperatures, of course, we cannot judge whether or not the
${\rm U}(1)_{\text{A}}$ symmetry is restored, since we lack the
lattice data in those temperatures. If we believe that the instanton
density is suppressed exponentially at high temperatures as is
expected by the Pisarski-Yaffe factor\cite{shuryak}, and the
correlation of the topological charges, i.e., $\chi$ is also
suppressed exponentially, the fitted line for the lattice data with a
Fermi function in Fig.\ \ref{fig3} (the dashed line) could be
considered as reasonable behavior. We notice here the deviations of
the NJL calculation (the solid line) from the fitted line at high
temperatures. As the CASE B, we let $K\Lambda^5$ have the temperature
dependence such that it reproduces the fitted line of $\chi$. The
calculated temperature dependence of $K\Lambda^5$ for this case is
shown in Fig.\ \ref{fig4}.

We notice that there are two lumps around $T_{\text{c}}$. It would be
senseless to take them seriously since we have ignored the temperature
dependence of $G\Lambda^2$ that should have been taken into account in 
principle. Rather, we should note that the ${\rm U}(1)_{\text{A}}$
symmetry is restored at high temperatures as is expected from the
starting assumption that the instanton density is suppressed at those
temperatures; the consistency is maintained in the NJL model.

We now calculate the constituent quark masses, the meson masses, and
the pion decay constant in our CASE A and CASE B. The results are
shown in Figs.\ \ref{fig5}$\sim$\ref{fig12}. The qualitative features
are almost the same as those of the CASE I by Hatsuda and
Kunihiro\cite{hk}.

Finally, we give the temperature dependence of $m_{\eta'}$ in
Figs.\ \ref{fig13} and \ref{fig14} that could be obtained by utilizing
the Witten-Veneziano mass formula (\ref{wv}). The $\eta'$ mass goes to
infinity at around $T_{\text{c}}=200\ {\rm MeV}$ in either CASE A or
B. This is because $f_\pi$ gets to zero at that temperature. We have
removed those infinities above the temperature at which $f_\pi$
vanishes because it is considered that our approximation scheme is
broken down there.

\section{Summary}
\label{sec4}
We have derived the expression for the topological susceptibility,
$\chi$, in the NJL model within the same approximation as for the
constituent quark masses and the meson masses, namely in the leading
order of large $N_{\text{c}}$ expansion. At zero temperature, we have
performed parameter setting by employing $\chi$ in place of
$m_{\eta'}$, and  have seen that the obtained parameters do not allow
the bound state of $\eta'$. At finite temperature, we have calculated
the temperature dependence of $\chi$, and have found that the lattice
data up to $1.4T_{\text{c}}$ are reproduced with a constant six-point
coupling constant $K$. This means that the ${\rm U}(1)_{\text{A}}$
symmetry is not restored up to $1.4T_{\text{c}}$, and we are led to
the conclusion that within the present framework, the effective
restoration of the ${\rm U}(1)_{\text{A}}$ symmetry and the chiral
phase transition do not necessarily occur simultaneously even though
the rapid dropping of $\chi$ around the chiral transition, observed in 
the lattice simulation, seemingly suggest the simultaneous
restoration. At high temperatures, we cannot state anything definitely
because of the absence of the lattice data. We have shown, however,
that if $\chi$ is suppressed exponentially, the
${\rm U}(1)_{\text{A}}$ symmetry is allowed to be restored at high
temperatures.

The topological susceptibility is an interesting quantity because it
is related to the mass of $\eta'$ through the Witten-Veneziano mass
formula. At zero temperature, we have seen that the formula is
satisfied numerically in the NJL model. At finite temperature, by
utilizing the formula, we have obtained knowledge as to the
temperature dependence of $m_{\eta'}$. In NJL model, $\eta'$ is far
from a stable particle even if it exists. Therefore we would say that
the parameter setting by using $m_{\eta'}$ is somewhat obscure. Our
approach proposed in this article is, on the other hand, not affected
by questionable quantities such as $m_{\eta'}$ and as a result it
takes the advantage to extract reliable results even when the bound
state of $\eta'$ cannot be available in the NJL model.

\section*{Acknowledgments}
We would like to thank T.\ Matsui, H.\ Fujii and M.\ Ohtani for
precious and valuable comments. One of authors (K.\ F.) thanks
L.\ von Smekal for the discussion about the interpretation of $\chi$
obtained here\cite{smekal}.

\begin{figure}
\caption{Three of various contracting ways are shown with their
corresponding diagrams.}
\label{fig1}
\end{figure}

\begin{figure}
\caption{Other leading order diagrams.}
\label{fig2}
\end{figure}

\begin{figure}
\caption{The lattice data are plotted with error bars. We have fitted
them with a Fermi function (the dashed line). The solid line denotes
the result of the NJL model with constant $K\Lambda^5$. }
\label{fig3}
\end{figure}

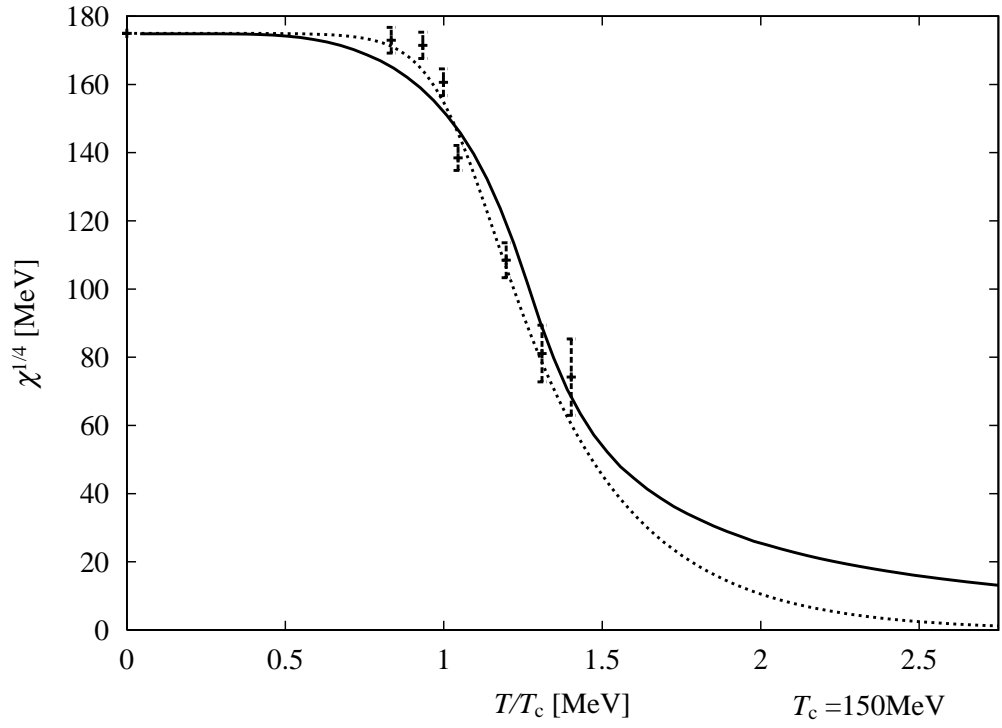
\begin{figure}
\caption{Temperature dependence of $K\Lambda^5$.}
\label{fig4}
\end{figure}

\begin{figure}
\caption{The constituent quark masses for CASE A.}
\label{fig5}
\end{figure}

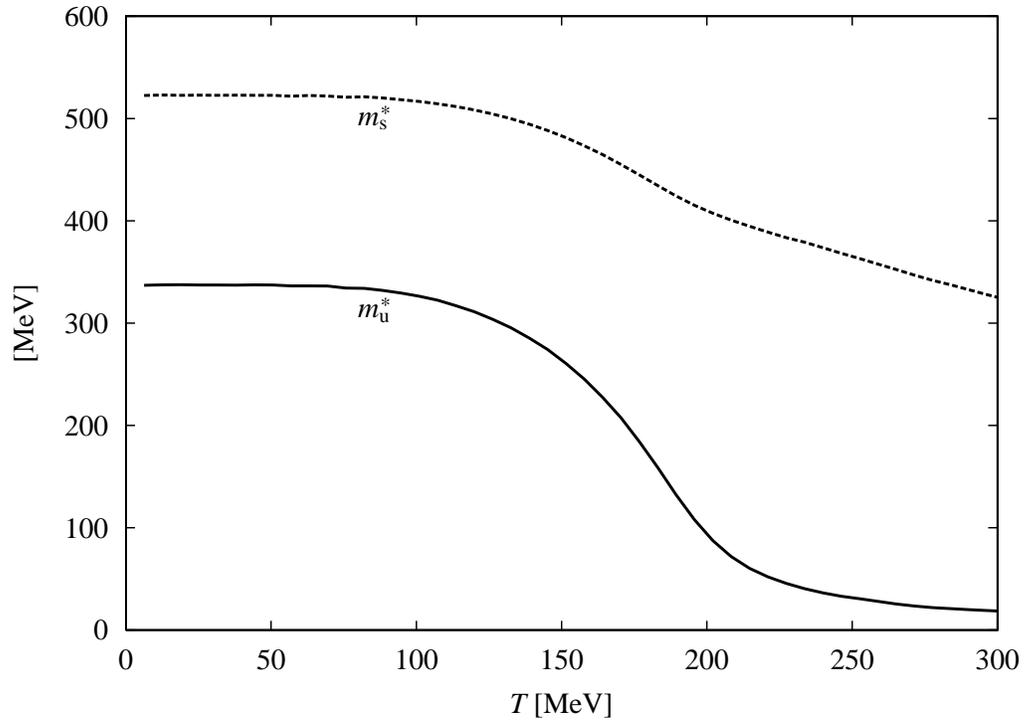
\begin{figure}
\caption{The constituent quark masses for CASE B.}
\label{fig6}
\end{figure}

\begin{figure}
\caption{The meson masses for CASE A.}
\label{fig7}
\end{figure}

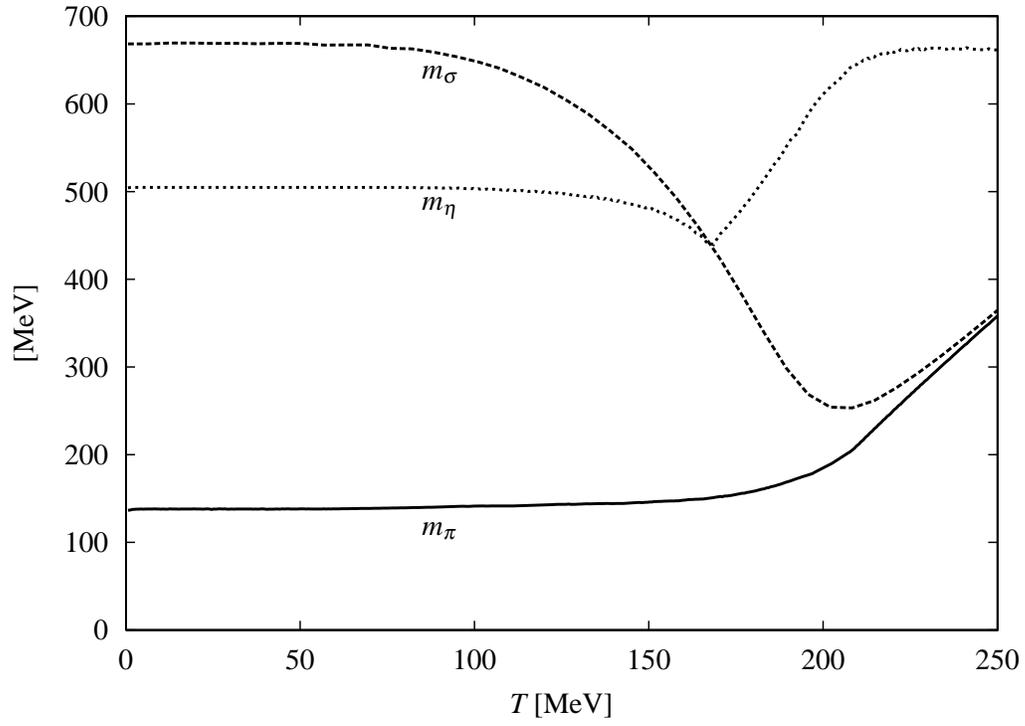
\begin{figure}
\caption{The meson masses for CASE B.}
\label{fig8}
\end{figure}

\begin{figure}
\caption{The Kaon mass for CASE A.}
\label{fig9}
\end{figure}

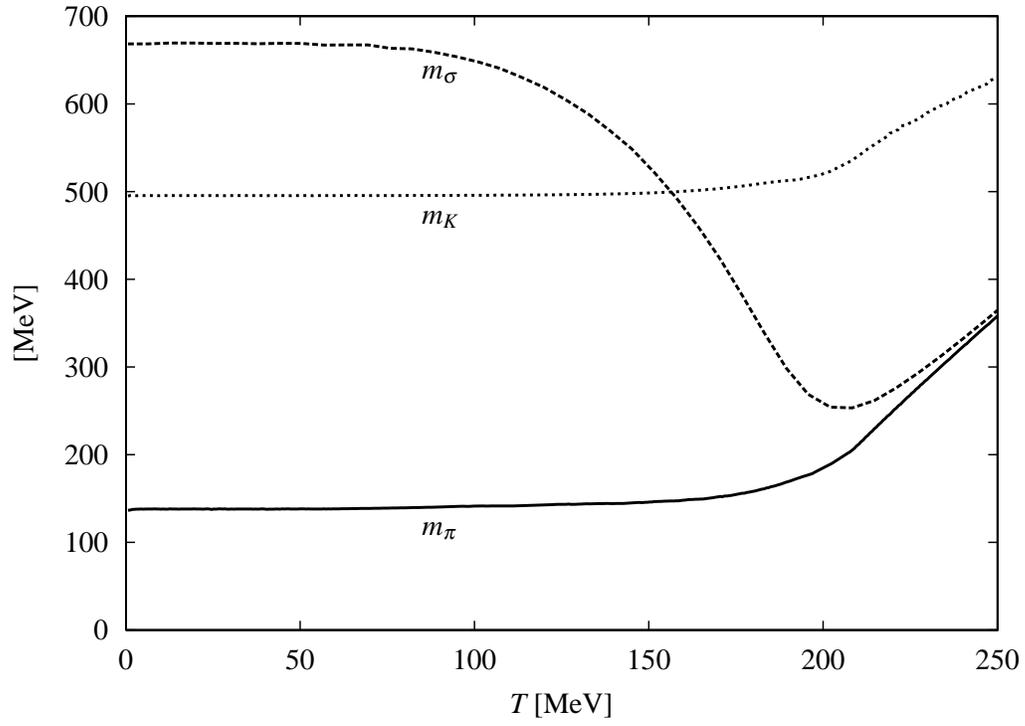
\begin{figure}
\caption{The Kaon mass for CASE B.}
\label{fig10}
\end{figure}

\begin{figure}
\caption{The pion decay constant for CASE A.}
\label{fig11}
\end{figure}

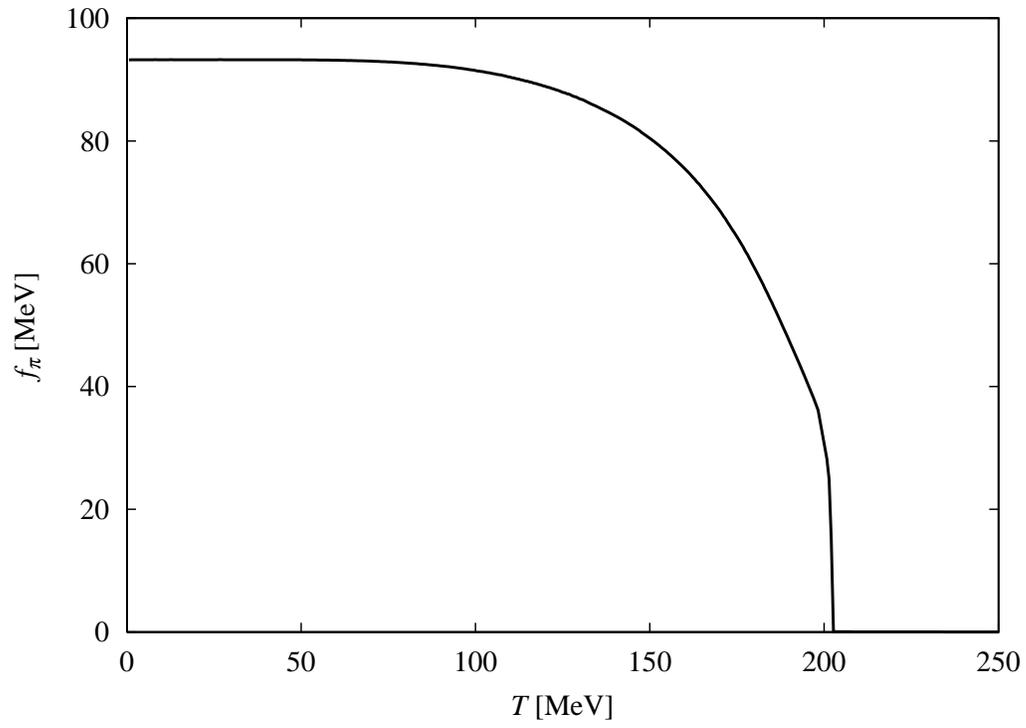
\begin{figure}
\caption{The pion decay constant for CASE B.}
\label{fig12}
\end{figure}

\begin{figure}
\caption{The $\eta'$ mass for CASE A.}
\label{fig13}
\end{figure}

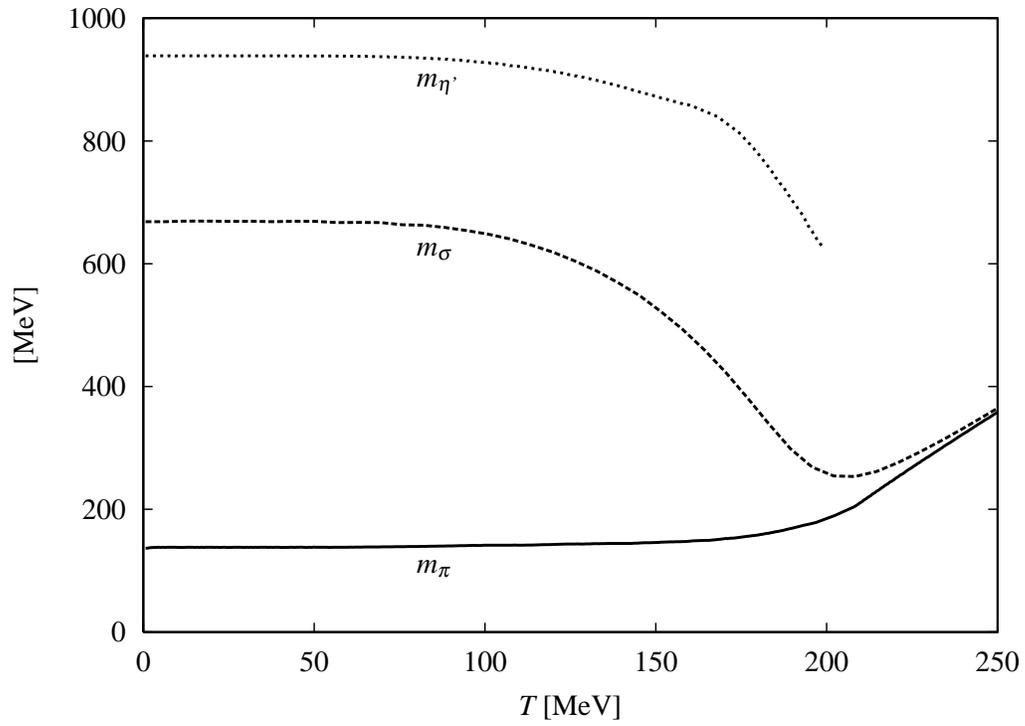
\begin{figure}
\caption{The $\eta'$ mass for CASE B.}
\label{fig14}
\end{figure}

\begin{table}
\begin{center}
\begin{tabular}{cccc}
\hline
 & Hatsuda and Kunihiro\cite{hk} & Ours
 & Experimental/empirical values \\
\hline
${m_{\text{u}}}^*$ (MeV) & 335   &   337   &   336 \\
${m_{\text{s}}}^*$ (MeV) & 527   &   523   &   540 \\
$m_\eta$ (MeV)           & 487   &   505   &   549 \\
$m_{\eta'}$ (MeV)        & (958) &   None [942]  &   958 \\
${\chi}^{1/4}$ (MeV)     & 166   &   (175) &   175 \\
$\theta_\eta$       & $-21^{\circ}$ & $-16.7^{\circ}$
 & $-20^{\circ}$ \\
\hline
\end{tabular}
\end{center}
\caption{Calculated physical quantities. Comparison of our results
with Hatsuda-Kunihiro and experimental data. In parentheses are the
values used as inputs. The value of $m_{\eta'}$ shown in the square
brackets is inferred by the Witten-Veneziano mass formula.}
\label{table}
\end{table}


\begin{references}
\bibitem{smekal}
L. von Smekal, in private communications; he pointed out that the
physical meaning of $\chi_{\text{NJL}}$ is closer to
$\chi_{\text{full}}$ rather than $\chi_{\text{pure}}$. We are now
making progress in our studies toward this direction.
\end{references}
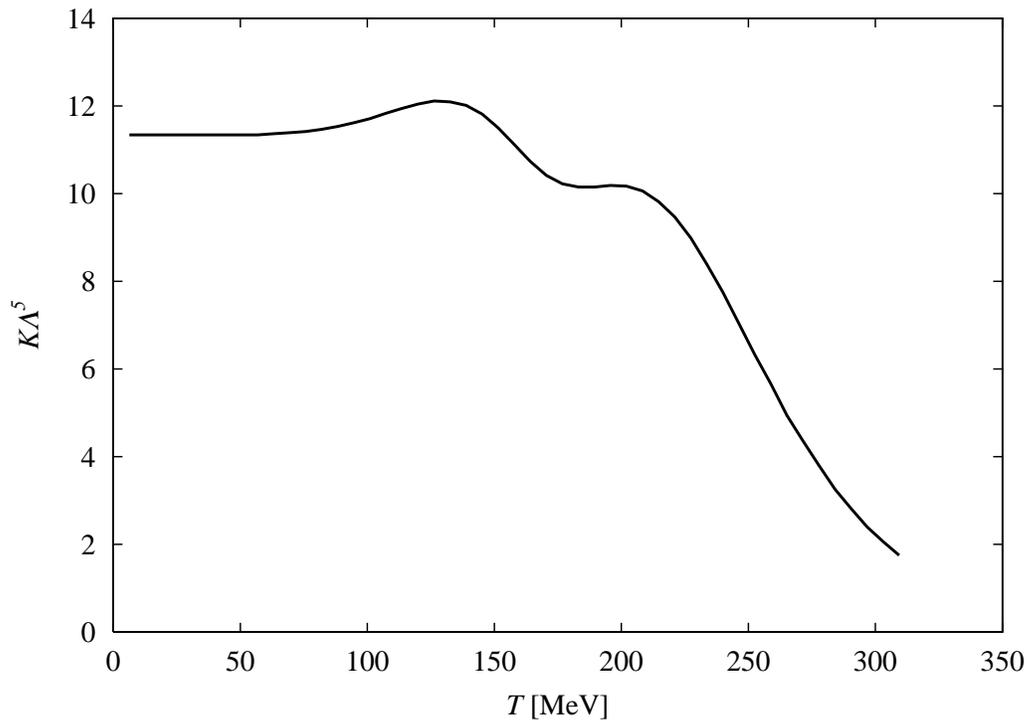
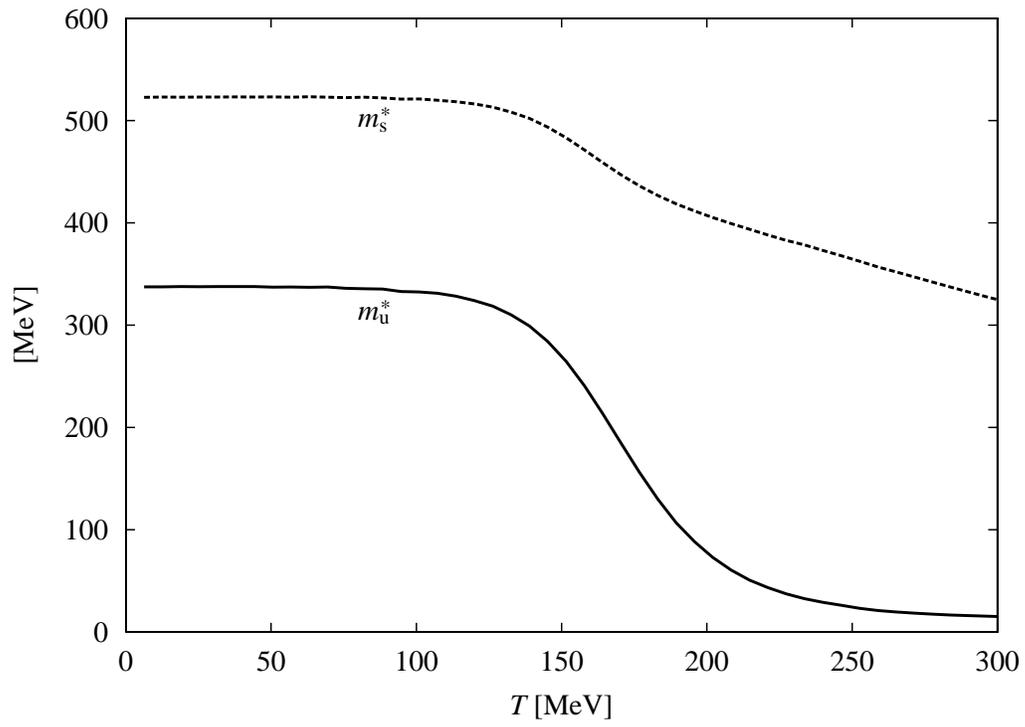
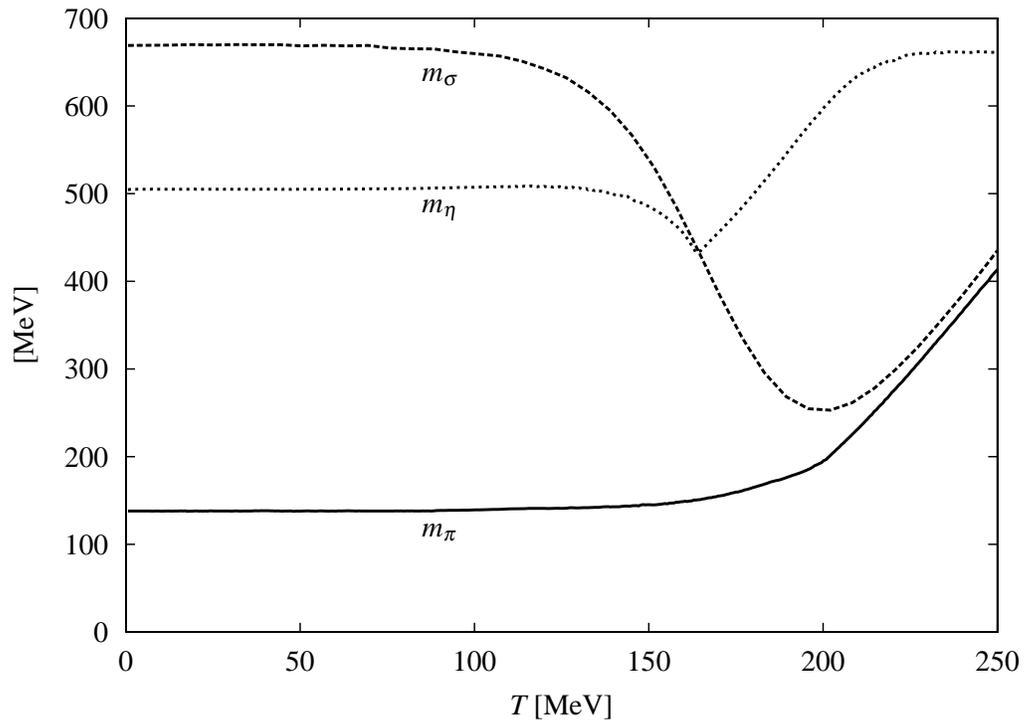
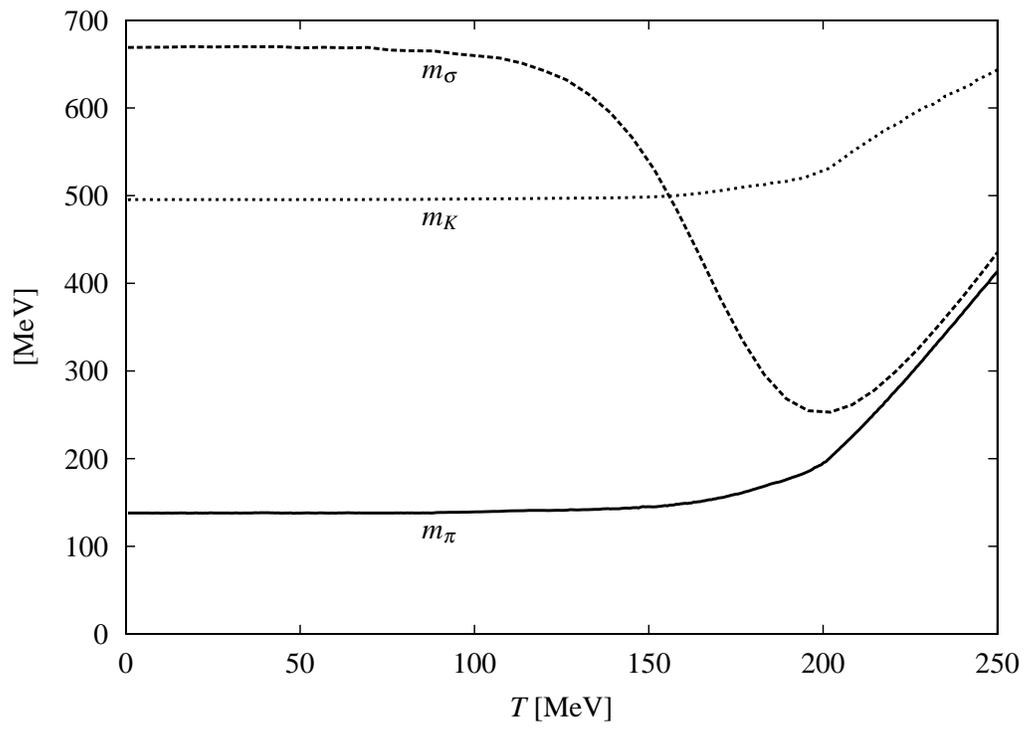
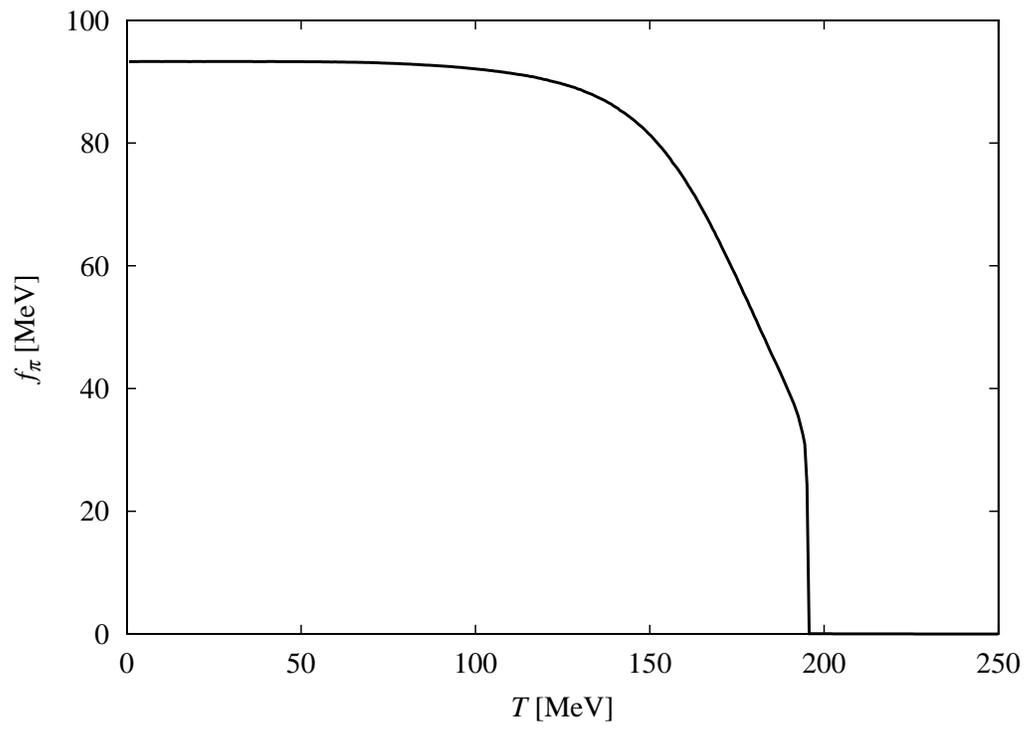
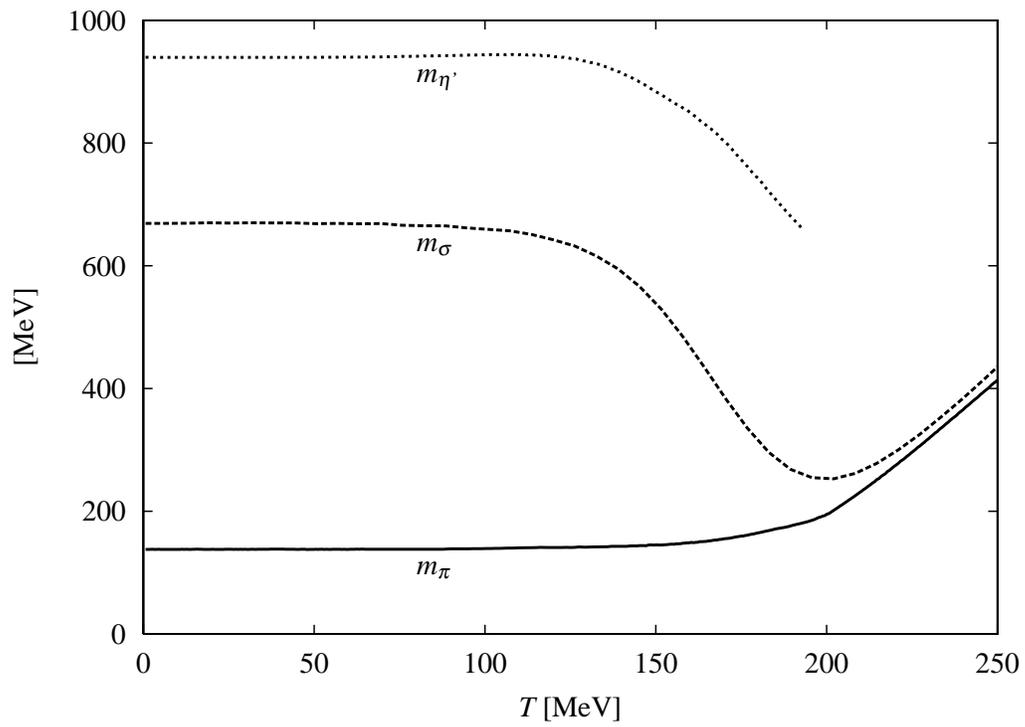
\end{document}